\documentclass[12pt,a4paper]{article}

\usepackage{amsmath,amssymb,graphicx,epsfig}

\begin{document}

\title{\bf Is thermo-ionic emission at room temperature exploitable?}

\author{Germano D'Abramo\vspace{0.2cm}\\
{\small Istituto Nazionale di Astrofisica,}\\ 
{\small Via Fosso del Cavaliere 100,}\\
{\small 00133, Roma, Italy.}\\
{\small E--mail: {\tt Germano.Dabramo@iasf-roma.inaf.it}}}

\vspace{0.2cm}
\date{\footnotesize{{\em Foundations of Physics}, under review since 
03--10--2008}. Revised version 07--05--2009.}

\maketitle

\begin{abstract}

In this brief note we describe two devices, a sort of flat and spherical 
capacitor, with which one should be able to test the possibility of 
creating a macroscopic voltage, and thus exploitable current, out of a 
single thermal source at room temperature. The basic idea is trivial and 
it makes use of a thermo-emitting cathode with work function as low as 
$0.7\,$eV. The idea is not completely new, but our approach is simpler and 
neat. When implemented, it should allow to assess if approaches based on 
thermo-ionic materials at room temperature really violate the Second Law 
of Thermodynamics macroscopically.\\ 


\end{abstract}

\section{Introduction}

The Second Law of Thermodynamics, explicitly in the form of Kelvin 
postulate, puts a fundamental limit to the way in which usable work can be 
extracted form heat sources, and to the maximum amount of work extractable 
(Carnot's principle). In particular, Kelvin postulate states that it is 
not possible to {\em cyclically} extract work from a {\em single} heat 
source. Clausius postulate of the Second Law, which is notoriously 
equivalent to Kelvin postulate, makes the impossibility more striking and 
understandable: heat can not spontaneously flow from sources at absolute 
temperature $T_1$ to sources at absolute temperature $T_2$ if $T_2\geq 
T_1$.

Although in the macroscopic physical world the Second Law seems 
authoritatively to make the difference between what is allowed and what is 
forbidden, in the microscopic world it seems to be continuously violated: 
let's take the Brownian motion or every fluctuation phenomena, for 
example.

As stated by Poincar\'e about Brownian motion~\cite{poin}:

\begin{quote}
``{\small [..]we see under our eyes now motion transformed into
heat by friction, now heat changed inversely into motion, and that without
loss since the movement lasts forever. This is contrary to the principle
of Carnot.}''.
\end{quote}

Almost every past attempt to understand and exploit such microscopic 
violation relies in the approach of {\em fluctuations rectification}. Even 
the famous thought experiment of {\em Maxwell's Demon} is based, in the 
end, on a idealized version of fluctuations rectification. The main 
difficulties with these approaches are that every macroscopic/microscopic 
rectifier device seems either to suffer fluctuations itself, which 
neutralize every usable net effect, or its functioning seems to increase 
the total entropy of the system, at least of the same amount of the 
alleged reduction, mainly because of energy dissipation and/or entropy 
cost in the acquisition of information needed to run the device. For a 
thorough historical account, see Earman and Norton~\cite{eanor}.

An approach to microscopic rectification that seems theoretically 
promising is the equivalent of the photo-electric effect with materials 
emitting electrons at room temperature. If we are able to collect even a 
fraction of the electrons emitted by these materials due to the absorption 
of radiation by a single heat source, then we would be able in principle 
to transform their flow into work and heat again, indefinitely. As a 
matter of fact, if {\em photo}-voltaic cells exist, why, {\bf in 
principle}, {\em thermo}-voltaic cells should not exist? As we will 
explain later in details, our approach appears not to suffer from the 
usual limitations~\cite{eanor}.

Nowadays thermo-ionic materials are available that have a non negligible 
electronic emission at room temperature. For example, the well known 
Ag--O--Cs photo-cathode has a low work function, $\phi\lesssim 0.7\,$eV, 
and a peak emission at about $\lambda=8500$\,\AA~\cite{somm1,somm2,bates}.

Other researchers have tried to experimentally implement similar ideas in 
the past (see, for example, \cite{shee}, \cite{fu1,fu2} and references 
therein). Unfortunately, their results, some of which have been published 
in peer-reviewed journals, have not received the attention they deserve 
from the scientific community.

As a matter of fact, those approaches share a common weakness: the current 
or the voltage produced are very small (microscopic in some cases) and, 
due to the complexity of the experimental apparatus (including the 
measurement equipment) and the complexity of the experimental conditions, 
it is often not possible to unambiguously decouple such a production from 
other sources known but not accounted for (magnetic fields, ordinary 
solid state and plasma physics effects, and so on).

In this note we provide a very simple and neat approach to test the 
exploitability of the `thermo-ionic emission idea'. In contrast with what 
has been previously done in our approach we do not need complex 
equipments, high temperatures, as in \cite{shee}, or any external, 
constant and uniform (or not uniform) magnetic field as rectifier device, 
as in \cite{fu1,fu2}. In such a way we should eliminate possible sources 
of spurious and not controllable electric effect, due for example to known 
thermo-electric phenomena, to the measurement equipment and to external 
magnetic fields. Moreover, the simplicity of our approach allows us to 
mathematically model its behavior with a straightforward analysis: in 
this way a comparison between theory and experimental results could be 
carried out on a solid, quantitative ground.

In a few words, what we describe in the following Sections is an analogous 
of a {\it thermo}-voltaic cell.

\section{Thermo-charged flat capacitor}

\begin{figure}[t]

\begin{center}
\begin{picture}(100,365)
\thicklines
\put(20,180){\oval(50,250)}
\put(-2,70){\line(0,1){220}}
\put(-1,70){\line(0,1){220}}
\put(0,70){\line(0,1){220}}
\put(41,70){\line(0,1){220}}
\put(-30,180){\line(1,0){30}}
\put(41,180){\line(1,0){30}}

\put(71,21){\line(0,1){160}}
\put(-30,21){\line(0,1){160}}
\put(41,270){\vector(1,1){30}}
\put(-1,270){\vector(-1,1){30}}
\put(70,305){{\scriptsize Metallic plate}}
\put(-80,305){{\scriptsize Plate with a}}
\put(-80,298){{\scriptsize thin layer of}}
\put(-80,291){{\scriptsize Ag--O--Cs on}}
\put(-80,284){{\scriptsize the internal side}}
\put(20,350){\vector(0,-1){60}}
\put(8,355){{\scriptsize Vacuum}}
\put(-1,200){\vector(1,1){20}}
\put(20,222){{\scriptsize $e^-$}}
\put(38,300){\vector(1,1){40}}
\put(52,345){{\scriptsize Opaque case}}
\put(20,180){\vector(-1,1){20}}
\put(22,175){{\scriptsize $h\nu$}}

\put(-30.5,21){\vector(1,0){35}}
\put(70.5,21){\vector(-1,0){35}}
\put(15.3,19){{\scriptsize $V$}}
\put(35,25){{\scriptsize $\ominus$}}
\put(0,25){{\scriptsize $\oplus$}}

\put(50,250){{\scriptsize Room temperature $T\simeq 298\,$K}}

\end{picture}
\end{center}
\caption{Scheme of the thermo-charged flat capacitor.}
\label{fig1}
\end{figure}
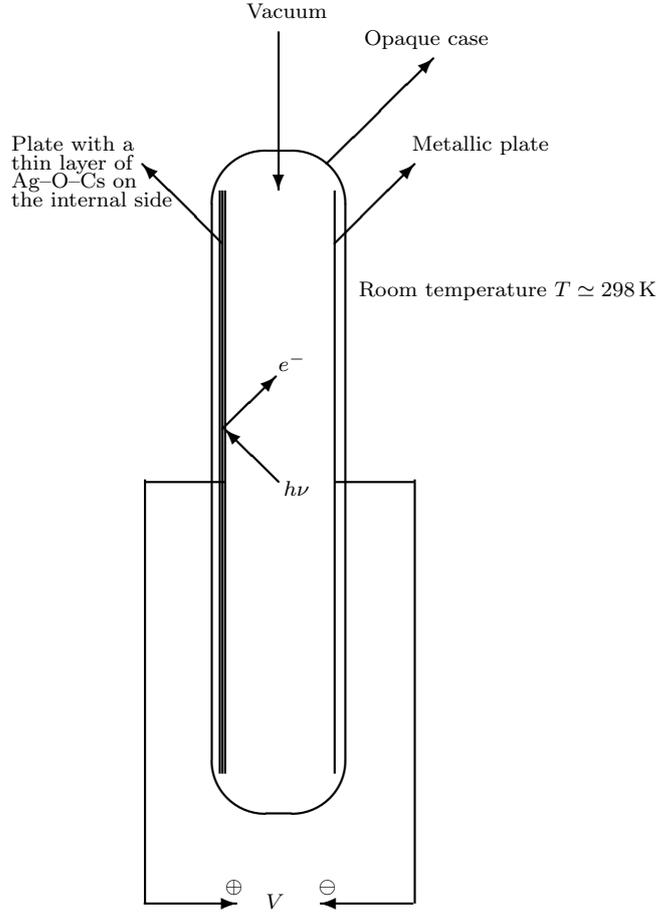

In Fig.~\ref{fig1} a sketch of our first device is shown. It is 
substantially a flat capacitor with square plates of area $S$ and 
inter-plates distance $d$, with $\sqrt{S}\gg d$ in order to reduce 
boundary effects. The capacitor should be placed inside a case that is 
opaque to the visible light (in order to avoid spurious {\em 
photo}-electric emission), under extreme vacuum and at room temperature. 
The {\em internal side} of one of the plates is covered with a thin layer 
of Ag--O--Cs material.

\begin{figure}[t]
\centerline{\includegraphics[width=6cm]{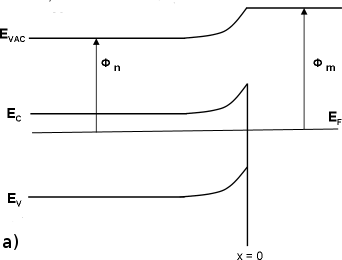}
  \includegraphics[width=6cm]{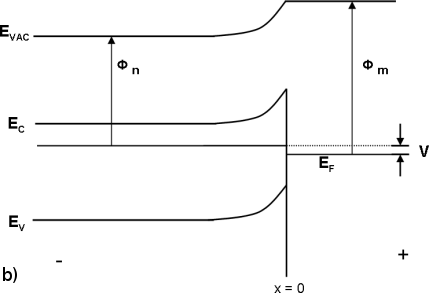}}
\caption{Figure a): Band profiles of semiconductor (n) - metal (m) junction at 
equilibrium. Figure b): Photo/thermo generated electrons accumulate in the 
n-type semiconductor. This rises the electron Fermi level and generates a 
further voltage $V$.}
\label{fig1b}
\end{figure}

Now, it should be clear how our device works. Almost half of the electrons 
emitted by the treated plate, due to thermo-ionic emission at room 
temperature, is collected by the second plate creating a macroscopic 
difference of potential $V$. Such process lasts until $V$ is 
too high to be overcome by the kinetic energy $K_e$ of the main fraction 
of emitted electrons (namely, when $K_e < eV$, where $e$ is the 
charge of electron). The other half of the emitted electrons, namely those 
emitted to the left in Fig.~\ref{fig1}, remains confined in the metallic 
structure of the treated plate. The contact surface between the metallic
plate and the Ag--O--Cs layer is a well known Schottky junction (metal/n-type
semiconductor). When two materials (in our case, a metal and a semiconductor)
are physically joined, so as to establish a uniform chemical potential, that 
is a single Fermi level, electrons are transferred from the material with the 
lesser work function $\phi_1$ (Ag--O--Cs) to the material with the greater
work function $\phi_2$ (metal). As a result a contact potential $V_c$ is 
established such that $eV_c=\phi_2-\phi_1$. The energy band profiles of
semiconductor-metal junction at equilibrium are shown in Fig.~\ref{fig1b}~a).
The aspect which counts for the functioning of our device is the fact that
the energy level of the vacuum for Ag--O--Cs (and that for the metallic 
plate) is preserved~\cite{ueb} far from the depletion region. This means that 
whenever an electron is extracted from the Ag--O--Cs to the vacuum 
(towards the second plate), and for what is said above this is made always at 
the cost of $\approx 0.7\,$eV, an electron must flow from the metallic plate 
to the Ag--O--Cs layer in order to re-establish the constant contact potential.
Thus, the contact potential is not a wall which forbids every flux of electrons 
through the junction metal/Ag--O--Cs. Moreover, thermal photons with energy 
greater or equal to the energy gap between the Fermi level and the conduction 
region in the semiconductor generate the accumulation of electrons in the 
semiconductor side, creating a further thermo-voltage across the 
junction: and all this is due to the contact field in the depletion region 
of the junction (see Fig.~\ref{fig1b}~b)).

As it will be clear later in the text, it is important to reduce to zero 
the amount of electrons which escape from the treated plate and, due to 
the open sides, are not collected by the second metallic plate. In the 
long run such a dissipation could lead to an exhaustion of the charging 
process. As a matter of fact, the condition $\sqrt{S}\gg d$ has been 
chosen also to maximally reduce such undesired effect. This problem, 
anyway, could be completely fixed using closed, spherical capacitors, as 
we will see in Section~3.

If a resistor of suitable ohmic resistance is placed between the plates, 
then a macroscopic and potentially exploitable current should be 
measurable.

It could be interesting to provide an estimate of the value of $V$ 
obtainable and an estimate of the time needed to reach such value, given 
the physical characteristics of the capacitor and the quantum efficiency 
curve $\eta(\nu)$ of the thermo-ionic material.

The capacitor is placed in a heat bath at room temperature and it is 
subject to the black-body radiation. Every plate, at thermal equilibrium, 
emits and absorbs an equal amount of radiation (Kirchhoff's law of thermal 
radiation), thus the amount of radiation absorbed by the thermo-ionic 
material is the same emitted by the plate according to the black-body 
radiation formula (Planck's equation). Given the room temperature $T$, 
Planck equation provides us with the number distribution of photons 
absorbed as a function of their frequency.

According to the law of thermo-ionic emission, the kinetic energy $K_e$ of 
the electron emitted by the material is given by the following equation,

\begin{equation}
K_e=h\nu-\phi,
\label{eq1}
\end{equation}
where $h\nu$ is the energy of the photon with frequency $\nu$ ($h$ is the 
Planck constant) and $\phi$ is the work function of the material. Thus, 
only the tail of the Planck distribution of photons absorbed, with frequency
$\nu>\nu_0=\phi/h$, can contribute to the thermo-ionic emission.

The voltage $V$ reachable with frequency $\nu_1$ is given by the 
following formula,

\begin{equation}
eV =h\nu_1-\phi,
\label{eq2}
\end{equation}
where $eV$ is the inter-plates potential energy, and thus,

\begin{equation}
\nu_1=\frac{eV +\phi}{h}.
\label{eq3}
\end{equation}

The total number of photons per unit time $F_{p}$, with energy greater 
than or equal to $h\nu_1$, emitted and absorbed in thermal equilibrium by 
the internal side of treated plate is given by the Planck equation,

\begin{equation}
F_{p}=\frac{2\pi S}{c^2}\int_{\nu_1}^\infty
\frac{\nu^2 d\nu}{e^{\frac{h\nu}{kT}}-1},
\label{eq4}
\end{equation}
where $S$ is the plate surface area, $c$ is the speed of light, $k$ is the 
Boltzmann constant and $T$ the room temperature.

If $\eta(\nu)$ is the quantum efficiency (or quantum yield) curve of the 
photo-cathode Ag--O--Cs, then the number of electrons per unit time $F_e$ 
with kinetic energy greater than or equal to $h\nu_1-\phi$ emitted (due to 
thermo-ionic effect) by the internal side of the treated plate towards the 
second metallic plate is given by,

\begin{equation}
F_{e}=\frac{2\pi S}{c^2}\int_{\nu_1}^\infty
\frac{\eta(\nu)\nu^2 d\nu}{e^{\frac{h\nu}{kT}}-1}.
\label{eq5}
\end{equation}

As it is well known, the electric charge $Q$ of a flat capacitor is linked 
to the voltage as follows,

\begin{equation}
V=\frac{d}{\epsilon_0 S}Q,
\label{eq6}
\end{equation}
where $\epsilon_0$ is the dielectric constant of vacuum.

Now, we derive the differential equation which governs the process of 
thermo-charging. In the interval of time $dt$ the charge collected by the 
metallic plate is given by,

\begin{equation}
dQ=eF_e dt=\frac{2\pi e S}{c^2}\Biggl(\int_{\frac{eV(t) +
\phi}{h}}^\infty \frac{\eta(\nu)\nu^2
d\nu}{e^{\frac{h\nu}{kT}}-1}\Biggr)dt, \label{eq7}
\end{equation}
where we make use of eq.~(\ref{eq3}) for $\nu_1$ and $V(t)$ is the 
voltage at time $t$. Thus, through the differential form of 
eq.~(\ref{eq6}), we have,

\begin{equation}
dV(t)= \frac{2\pi e d}{\epsilon_0
c^2}\Biggl(\int_{\frac{eV(t) + \phi}{h}}^\infty
\frac{\eta(\nu)\nu^2 d\nu}{e^{\frac{h\nu}{kT}}-1}\Biggr)dt,
\label{eq8}
\end{equation}
or

\begin{equation}
\frac{dV(t)}{dt}= \frac{2\pi e d}{\epsilon_0
c^2}\int_{\frac{eV(t) + \phi}{h}}^\infty \frac{\eta(\nu)\nu^2
d\nu}{e^{\frac{h\nu}{kT}}-1}.
\label{eq9}
\end{equation}

Unfortunately, provided that an analytical approximation of a real quantum 
efficiency curve $\eta(\nu)$ exists, the previous differential equation 
appears to have no general, simple analytical solution.

However, a close look at the Planckian integral of eq.~(\ref{eq9}) 
suggests us the asymptotic behavior of $V(t)$. Even if we do not 
know {\it a priori} how $\eta(\nu)$ is, we know it to be a bounded 
function of frequency, with values between zero and one; usually, the 
higher is $\nu$, the closer to $1$ is $\eta(\nu)$. Thus, independently 
from $\eta(\nu)$, a slight increase of $V(t)$ makes the value of 
the Planckian integral to be smaller and smaller very fast. Heuristically, 
this suggests that $V(t)$ should tend quite rapidly to an 
`asymptotic' value (since $\frac{dV}{dt}$ tends to $0$).

In the rest of this section we provide a numerical solution of the above 
differential equation. To do so we need to adopt an approximation, 
however: the approximation we make consists in the adoption of a constant 
value for $\eta$, a sort of suitable mean value $\overline{\eta}$.

The differential equation~(\ref{eq9}) thus becomes,

\begin{equation}
\frac{dV(t)}{dt}= \frac{2\pi e d\overline{\eta}}{\epsilon_0
c^2}\int_{\frac{eV(t) + \phi}{h}}^\infty \frac{\nu^2
d\nu}{e^{\frac{h\nu}{kT}}-1}.
\label{eq10}
\end{equation}

A straightforward variable substitution in the integral of 
eq.~(\ref{eq10}) allows us to write it in its final simplified form,

\begin{equation}
\frac{dV(t)}{dt}= \frac{2\pi e d\overline{\eta}}{\epsilon_0
c^2}\biggl(\frac{kT}{h}\biggr)^3\int_{\frac{eV(t) +
\phi}{kT}}^\infty \frac{x^2 dx}{e^x-1}.
\label{eq11}
\end{equation}

\begin{figure}[t]
\centerline{\includegraphics[width=7cm, height=4.5cm]{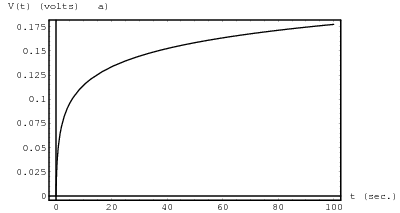}
  \includegraphics[width=7cm, height=4.5cm]{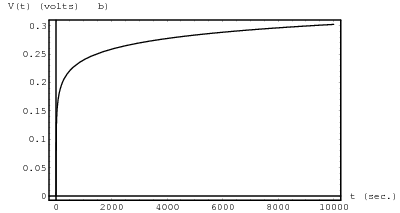}}
\caption{Thermo-charging process of the flat capacitor described in the text
($\phi=0.7\,$eV, $d=0.05\,$m, $T=298\,$K, and $\overline{\eta}=10^{-5}$).
These two plots show with different ranges in time-scale the behavior of
$V(t)$. Plot a) shows how only after 60 seconds the voltage of the
capacitor becomes more than $0.15\,$volts. Instead, plot b) tells us that
the voltage of the capacitor requires hundreds of hours to approach
$0.3\,$volts.}
\label{fig2}
\end{figure}

Here we provide an exemplificative numerical solution of eq.~(\ref{eq11}), 
adopting the following nominal values for $\phi$, $d$, $T$ and 
$\overline{\eta}$: $\phi=0.7\,$eV, $d=0.05\,$m, $T=298\,$K, and 
$\overline{\eta}=10^{-5}$. In order to make a conservative choice for the 
value of $\overline{\eta}$ we note that only black body radiation with 
frequency greater than $\nu_0=\phi/h$ can contribute to the thermo-ionic 
emission. This means that for the Ag--O--Cs photo-cathode only radiation 
with wavelength smaller than $\lambda_0=hc/\phi\simeq 1700\,$nm 
contributes to the emission. According to Fig.~1 in Bates~\cite{bates}, 
the quantum yield of Ag--O--Cs for wavelengths smaller than $\lambda_0$ 
(and thus, for frequency greater than $\nu_0$) is always greater than 
$10^{-5}$.  Anyway, a laboratory realization of the capacitor, together 
with the experimental measurement of $V(t)$, should provide us with 
a realistic estimate of $\overline{\eta}$ for the photo-cathode Ag--O--Cs.

\begin{figure}[t]
\centerline{\includegraphics[width=7cm, height=4.5cm]{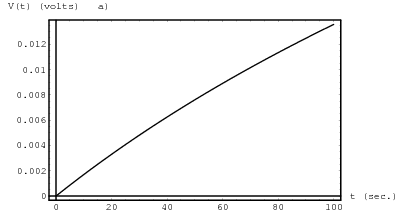}
  \includegraphics[width=7cm, height=4.5cm]{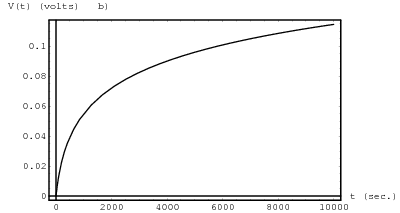}}
\caption{Thermo-charging process of the flat capacitor described in the text
with $\phi=0.7\,$eV, $d=0.05\,$m, $T=298\,$K, and
$\overline{\eta}=10^{-8}$. These two plots show with different ranges in
time-scale the behavior of $V(t)$. Plot a) shows how only after 60
seconds the voltage of the capacitor is near to $0.01\,$volts. Instead,
plot b) tells us that the voltage of the capacitor requires hundreds of
hours to become equal to $0.1\,$volts.}
\label{fig3}
\end{figure}

In Fig.~\ref{fig2} the numerical solution of the above test it is shown. 
In plot a) we could easily see how only after 60 seconds the voltage of 
the capacitor exceeds the value of $0.15\,$volts. Indeed, this is a 
macroscopic voltage, and it is exploitable as well since it requires a 
relatively short charging time. Instead, plot b) tells us that the voltage 
of the capacitor requires hundreds of hours to approach $0.3\,$volts. Even 
in the more pessimistic scenario where $\overline{\eta}=10^{-8}$ we see 
that a macroscopic voltage should arise quite rapidly between the plates, 
see Fig.~\ref{fig3}.

\section{Thermo-charged spherical capacitor}

Although the geometrical condition $\sqrt{S}\gg d$ chosen for the flat 
capacitor greatly reduces the electrons loss due to the open sides of the 
device, such undesired effect persists anyway and in the long run leads to 
an exhaustion of the charging process (ending with both plates having the 
same positive charge).

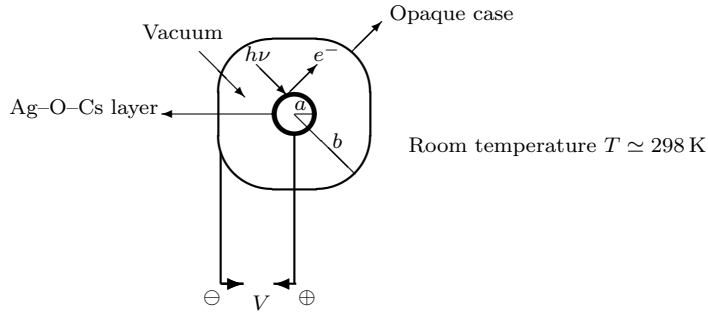
\begin{figure}[t]

\begin{center}
\begin{picture}(150,150)
\setlength{\unitlength}{0.5cm} \thicklines \put(4,6){\circle{1}}
\put(4,6){\circle{1.08}} \put(4,6){\circle{1.1}}
\put(4,6){\oval(4,4)}
\put(4,5.5){\line(0,-1){4}}\put(4.05,1.5){\vector(-1,0){0.6}}
\put(2.06,1.5){\vector(1,0){0.6}} \put(2.06,5){\line(0,-1){3.5}}
\put(2.9,0.8){{\scriptsize $V$}} \put(1.6,1){{\scriptsize
$\ominus$}} \put(4.1,1){{\scriptsize $\oplus$}}
\put(6.5,8.5){{\scriptsize Opaque case}}
\put(2.06,1.5){\vector(1,0){0.6}} \put(0,8){{\scriptsize Vacuum}}
{\thinlines\put(5.5,7.65){\vector(1,1){0.8}}
\put(1.5,7.65){\vector(1,-1){1.2}}} \put(4,6.1){{\scriptsize $a$}}
\put(5,5.1){{\scriptsize $b$}}
{\thinlines\put(4,6){\line(1,0){0.5}}}
{\thinlines\put(4,6){\line(1,-1){1.6}}} \put(7,5){{\scriptsize Room
temperature $T\simeq 298\,$K}} {\thinlines
\put(3.5,6){\vector(-1,0){3}}} \put(-3.5,6){{\scriptsize Ag--O--Cs
layer}} {\thinlines \put(3,7.3){\vector(1,-1){0.8}}} {\thinlines
\put(3.8,6.5){\vector(1,1){0.8}}} \put(2.7,7.4){\scriptsize $h\nu$}
\put(4.5,7.4){\scriptsize $e^-$}


\end{picture}
\end{center}
\caption{Scheme of the thermo-charged spherical capacitor.}
\label{fig4}
\end{figure}

In this Section we provide an analysis, similar to that done before, for a 
closed, spherical capacitor. Such device does not suffer from electrons 
dispersion. In Fig.~\ref{fig4} a sketched section of the spherical 
capacitor is shown. The outer sphere has radius $b$, while the inner one 
has radius $a$. The outer surface of the inner sphere is covered with a 
thin layer of Ag--O--Cs. As in the previous Section, the number of 
electrons per unit time $F_e$ with kinetic energy greater than or equal to 
$h\nu_1 -\phi$ emitted by the outer surface of the inner sphere towards 
the outer one is given by,

\begin{equation}
F_{e}=\frac{2\pi\cdot 4\pi a^2}{c^2}\int_{\nu_1}^\infty
\frac{\eta(\nu)\nu^2 d\nu}{e^{\frac{h\nu}{kT}}-1},
\label{eq12}
\end{equation}
where $4\pi a^2$ is the surface area of the inner sphere.

For a spherical capacitor, the voltage $V$ between the spheres and 
the charge $Q$ on each are linked by the following equation,

\begin{equation}
V=\frac{Q}{4\pi \epsilon_0}\frac{b-a}{ab}.
\label{eq13}
\end{equation}

In order to derive the differential equation which governs the charging 
process for the spherical capacitor we proceed as in the previous Section 
for eq.~(\ref{eq9}), obtaining the following equation,

\begin{equation}
\frac{dV(t)}{dt}= \frac{2\pi e}{\epsilon_0
c^2}\frac{a(b-a)}{b} \int_{\frac{eV(t) + \phi}{h}}^\infty
\frac{\eta(\nu)\nu^2 d\nu}{e^{\frac{h\nu}{kT}}-1}.
\label{eq14}
\end{equation}

Since our aim is to maximize the production of $V$, we have to 
choose $a$ and $b$ such that they maximize the geometrical factor 
$a(b-a)/b$. It is not difficult to see that the maximum is reached when 
$a=b/2$. So we have,

\begin{equation}
\frac{dV(t)}{dt}= \frac{\pi e b}{2\epsilon_0
c^2}\int_{\frac{eV(t) + \phi}{h}}^\infty \frac{\eta(\nu)\nu^2
d\nu}{e^{\frac{h\nu}{kT}}-1}.
\label{eq15}
\end{equation}

Applying the same approximation for $\eta(\nu)$ and the variable 
substitution used for eq.~(\ref{eq11}), we finally have,

\begin{equation}
\frac{dV(t)}{dt}= \frac{\pi e b\overline{\eta}}{2\epsilon_0
c^2}\biggl(\frac{kT}{h}\biggr)^3\int_{\frac{eV(t) +
\phi}{kT}}^\infty \frac{x^2 dx}{e^{x}-1}.
\label{eq16}
\end{equation}

We note that if we built a spherical capacitor with $b=4d$, where $d$ is 
the inter plates distance of the previous flat capacitor, then its 
charging process is governed by the very same equation of Section~2, with 
the same solutions shown in Fig.~\ref{fig2} and Fig.~\ref{fig3}.

\section{Concluding remarks}

In this paper we have presented two sort of `thermo-charged' capacitors: 
they are a flat and a spherical vacuum capacitor with one conductor coated 
with thermo-emitting material Ag--O--Cs. We have also mathematically 
modeled their behavior and the results show that our devices are 
theoretically promising (other than being of simple construction) as a 
tool for testing the validity of the Second Law of Thermodynamics on a 
macroscopic level. If the experimental trend of $V$ with time were 
well fitted by eq.~(\ref{eq11}) (or eq.~(\ref{eq16})) with a suitable 
values of $\overline{\eta}$ and $\phi$, then this would mean that our 
mathematical model describes quite well the actual physical process of 
thermo-charging and also that the Second Law of Thermodynamics is under 
threat with fewer ambiguities.

Finally, we stress that an experimental verification of the expected 
functioning of the capacitors urges, since what is at stake is one of the 
most important law of Nature, with valuable and profound consequences.

\end{document}